\documentclass[twocolumn, pra, showpacs, aps, superscriptaddress]{revtex4-1}

\usepackage{amsmath}
\usepackage{amsfonts}
\usepackage{bbold}

\usepackage{graphicx}
\usepackage{ae,aecompl}
\usepackage{bm}
\usepackage{textcomp}
\usepackage{xcolor}

\usepackage{wasysym}
\usepackage{hyperref}

\begin{document}


\date{\today}
\title{Bichromatic force on metastable Argon for Atom Trap Trace Analysis}

\author{Z. Feng}
\email{bcf@matterwave.de}
\author{S. Ebser}
\author{L. Ringena}
\author{F. Ritterbusch}
\author{M.~K. Oberthaler}

\affiliation{Kirchhoff-Institute for Physics, Heidelberg University, Im Neuenheimer Feld 227, 69120 Heidelberg, Germany.}


\begin{abstract}
For an efficient performance of Atom Trap Trace Analysis, it is important to collimate the particles emitted from an effusive source. Their high velocity limits the interaction time with the cooling laser. Therefore forces beyond the limits of the scattering force are desirable. The bichromatic force is a promising candidate for this purpose which is demonstrated here for the first time on metastable argon-40. The precollimated atoms are deflected in one dimension and the acquired Doppler shift is detected by absorption spectroscopy. With the experimentally accessible parameters, it was possible to measure a force three times stronger than the scattering force. Systematic studies on its dependence on Rabi frequency, phase difference and detuning to atomic resonance are compared to the solution of the optical Bloch equations. We anticipate predictions for a possible application in Atom Trap Trace Analysis of argon-39 and other noble gas experiments, where a high flux of metastable atoms is needed.
\end{abstract}
\maketitle 

\section{Introduction} 
The isotopes of noble gases are of great importance for radiometric dating as they are chemically inert and a part of the atmospheric composition. Together with $^{14}$C and $^{85}\text{Kr}$, the long-lived $^{39}$Ar and $^{81}$Kr cover dating ranges from present to a million years ago \cite{Loosli2000}, but the low abundance $^{39}\text{Ar/Ar} \approx 8 \times 10^{-16}$ and $^{81}\text{Kr/Kr} \approx 5 \times 10^{-13}$ hinders routine measurements with established techniques such as Low-Level Counting (LLC) and Accelerator Mass Spectrometry (AMS) due to the limited sample size, measurement time and costs \cite{Lu2014}.

Atom Trap Trace Analysis (ATTA) exploits the shift of the optical resonance frequency due to differences in mass and nuclear spin to detect these rare isotopes down to the part-per-quadrillion level \cite{Lu2014}. Although a single resonant excitation is not sufficient to distinguish between the radioisotope and the huge background of abundant isotopes, many cycles of photon absorption and spontaneous emission in a Magneto-Optical Trap (MOT) strongly enhance the sensitivity and guarantee perfect selectivity. This technique is already in use for routine measurements of $^{85}$Kr and $^{81}$Kr \cite{Chen1999, Jiang2012}. The method has been further demonstrated with $^{39}$Ar \cite{Jiang2011}. A stable atmospheric count rate of $[3.58 \pm 0.10] \text{\,atoms}/\text{h}$ has been achieved for the first measurement of the $^{39}$Ar concentration in environmental samples using ATTA \cite{Ritterbusch2014}.

The low count rate is the most limiting factor for $^{39}$Ar ATTA and increasing the metastable atom flux is of utmost importance for the broad applications of this technique. Collimation of the effusive source is a crucial part and is addressed by an arrangement of tilted beams acting via the spontaneous scattering force (see Fig.\,\ref{fig:interactionzone}). This has increased the capture efficiency by two orders of magnitude, but still, only about 6\,\% of the total metastable atoms are collimated. A precollimation setup (transverse kicker) for slowing down faster atoms has therefore great potential for further improvement of the count rate. However, due to the short interaction time and the saturation of the scattering force, stronger mechanisms are desirable. 

The bichromatic force is a candidate for this purpose and has been shown for Cs \cite{Soding1997}, Na \cite{Voitsekhovich1994}, Rb \cite{Williams1999} and metastable helium \cite{Cashen2001}. This work demonstrates the bichromatic force for metastable argon. It has been measured for the abundant $^{40}$Ar here, as systematic measurements on $^{39}$Ar are not feasible at the moment.

\section{Theoretical description} 
The bichromatic force arises in counterpropagating laser beams consisting of two frequencies detuned by $\pm \Delta$ from the atomic resonance $\omega_{A}$. It can be considered as a rectification of the dipole force to create a non-vanishing average force which is, unlike the scattering force, not fundamentally limited by the scattering rate and can be made much larger than the latter if detuning $\Delta$, relative phase $\phi$ and intensity $I$ are chosen properly \cite{Metcalf1999}.

\begin{figure*}[t!] 
\includegraphics[width = 178mm]{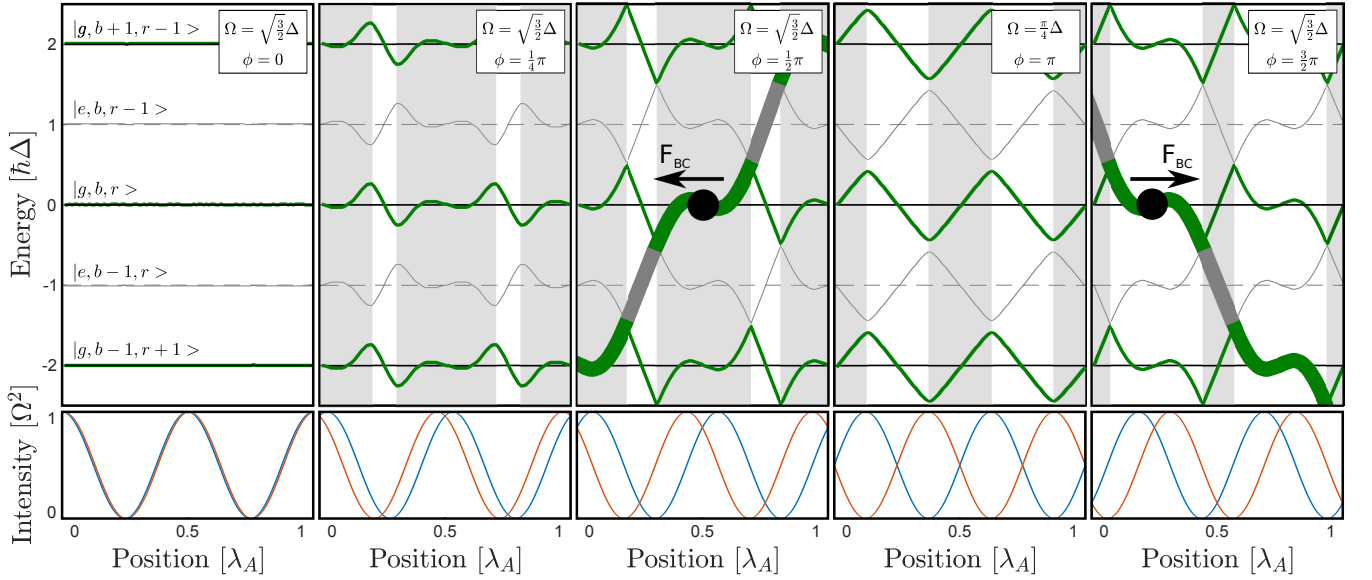}
\caption{(color online). {\bf Position-dependent energy levels of the system for different phases of the standing waves.} The horizontal lines mark the uncoupled ground (solid) and excited (dashed) states. An atom will undergo transitions between the dressed states with a certain probability and will gain or lose kinetic energy in this process. The preferred direction of the bichromatic force is given by the position and the phase difference $\phi$ of the standing waves. For $\phi = 0$, the coupling to both modes compensate each other and the states remain uncoupled. In the case of $\phi = \frac{1}{2}\pi$, atoms which start in the shaded area of a state with larger admixture of the ground state (bold curves), will receive a force to the left side. The opposite force acts on atoms starting in the unshaded area, but due to the three-to-one ratio of the shaded area, the mean force is to the left. For $\phi = \frac{3}{4}\pi$, the sign is reversed and the average force is to the right. For $\phi = \pi$, the optimal Rabi frequency is $\Omega = \frac{\pi}{4}\Delta$, but no direction is favored. Other choices of phase difference yield non-optimal results, since the level gaps are increased. Spontaneous emission occurs more likely in the states with larger admixture to the excited state (gray curves) and will change the slope direction. However, the force is retained on average as a subsequent spontaneous emission is more likely to happen once the path has changed.} 
\label{fig:dda}
\end{figure*}

An analytical approach is given by the doubly-dressed atom (DDA) model first proposed in \cite{Yatsenko2004}. The bare atom states are $|g\rangle$ and $|e\rangle$ and the light field is denoted by $|b,r\rangle$, where $b$ and $r$ are the number of blue and red detuned photons. By neglecting interactions with the light fields, the undressed product states
\begin{equation} \label{eq:undressedstates}
	| g, b, r \rangle = | g \rangle \otimes | b, r \rangle \text{ and } | e, b, r \rangle = |e \rangle \otimes | b, r \rangle
\end{equation}
form an infinite ladder with eigenenergies $E_{e, b, r} = b \cdot \hbar(\omega_{A}+\Delta) + r \cdot \hbar(\omega_{A}-\Delta) + \hbar \omega_{A} = E_{g, b, r} + \hbar \omega_{A}$ as in the ordinary dressed atom picture. In contrast to the case of one frequency, each level with conserved total number of photons $n = b + r$ is a manifold by itself separated by $\text{d}E = 2 \hbar \Delta$ due to splitting into the two light modes.

The Hamiltonian $\hat{H}_{NC}$ of the uncoupled system is the sum of an atomic part $\hat{H}_{0} = \hbar \omega_{A} | e \rangle \langle e |$ and light fields $\hat{H}_{b} + \hat{H}_{r} = \hbar(\omega_{A} + \Delta) \hat{a}^{\dagger}_{b} \hat{a}_{b} + \hbar(\omega_{A}-\Delta)\hat{a}^{\dagger}_{r} \hat{a}_{r}$, where $\hat{a}^{\dagger}_{b,r}$ and $\hat{a}_{b,r}$ are the photon creation and annihilation operators acting on $| b , r \rangle$. The interaction term
	\begin{equation}
		\begin{aligned}
		\hat{H}_{int} &= \frac{\hbar\Omega_{b}}{2}\left( \hat{a}^{\dagger}_{b} |g\rangle \langle e| + \hat{a}_{b} |e\rangle \langle g|\right) \\
		&+ \frac{\hbar\Omega_{r}}{2}\left( \hat{a}^{\dagger}_{r} |g\rangle \langle e| + \hat{a}_{r} |e\rangle \langle g|\right)
		\end{aligned}
	\end{equation}
can be described by introducing coupling parameters $\Omega_{b}$ to the blue and $\Omega_{r}$ to the red light field. 

In this model, the occupation of the bichromatic field modes can be coherently redistributed by the atom. For absorption of a red detuned photon with $E_{r} = \hbar (\omega_{A} - \Delta)$ and stimulated emission of a blue detuned photon $E_{b} = \hbar (\omega_{A} + \Delta)$, the light field gains energy of $\text{d}E = 2 \hbar\Delta$ and the atom loses this amount of kinetic energy. This is the origin of the bichromatic force. Changing the order of red and blue detuned photons will reverse its sign.

The coupling parameters are given by the choice of the light field configuration. In the case of two standing waves detuned by $\pm \Delta$ with relative phase difference $\phi$, they result in the space dependent term \cite{Partlow2004}
	\begin{equation} \label{eq:doubledressedrabi}
		\Omega_{b,r}(z) = 2 \Omega \cos(kz \pm \frac{\phi}{4})
	\end{equation}
with wavenumber $k = 2\pi/\lambda_{A}$ and position $z$. The Rabi frequency $\Omega = \gamma \sqrt{\frac{I}{2 I_{S}}}$ is given by intensity $I$, saturation intensity $I_{S}$ and natural linewidth $\gamma$. Truncated to a $5 \times 5$ matrix, the total Hamiltonian
	\begin{equation} \label{eq:dressedstateshamiltonian}
		\hat{H}_{tot} = (b+r) \hbar \omega_{A} \cdot \mathbb{1}_{5} + \hbar
		\begin{pmatrix}
			2\Delta & \frac{\Omega_{b}}{2} & 0 & 0 & 0\\
			\frac{\Omega_{b}}{2} & \Delta & \frac{\Omega_{r}}{2} & 0 & 0 \\
			0 & \frac{\Omega_{r}}{2} & 0 & \frac{\Omega_{b}}{2} & 0 \\
			0 & 0 & \frac{\Omega_{b}}{2} & -\Delta & \frac{\Omega_{r}}{2}\\
			0 & 0 & 0 & \frac{\Omega_{r}}{2} & -2\Delta \\
		\end{pmatrix}
	\end{equation}
is then also position-dependent and can be written in matrix form using the uncoupled states as basis. 

An atom with non-zero velocity $v$ will perform Landau-Zener transitions with a probability $P_{\text{LZ}} = \exp\left(-\pi U^{2}/\hbar v \frac{\text{d}E}{\text{d}z} \right)$, where $U$ is the energy gap between the states and $\frac{\text{d}E}{\text{d}z}$ the gradient of those. 
The maximal probability and therefore the maximal force is obtained when crossings of the energy levels occur. This is fulfilled by choosing $\Omega = \sqrt{\frac{3}{2}}\Delta$ and $\phi = \pm \frac{\pi}{2}$  (see Fig.\,\ref{fig:dda}). The system will then follow a path through the dressed states while the atom gains or loses kinetic energy by climbing up or sliding down the slopes \cite{Metcalf1999}. The preferred direction is determined by the starting position of the groundstate atoms and yields a three-to-one ratio for energy loss versus gain. The average bichromatic force in the doubly-dressed atom model is then
	\begin{equation} \label{eq:dressedstateforce}
		F_{\text{DDA}} = \frac{3}{4} \left| \frac{\text{d}E}{\text{d}z}  \right| - \frac{1}{4} \left| \frac{\text{d}E}{\text{d}z}  \right| = \frac{\hbar k \Delta}{\pi} \text{,}
	\end{equation}
as three-quarter of the atoms will follow the desired path and one-quarter are going in the other direction \cite{Corder2015}. This coherent process is disturbed by incoherent spontaneous emissions, which have to be added to the description. They will change the direction of the followed path and occur more likely in levels which have larger admixture of the uncoupled excited states. However, due to the choice of an asymmetric phase difference $\phi = \pm \frac{\pi}{2}$, the force in the desired direction is restored as a subsequent spontaneous emission is more likely to happen once the path has changed.

\begin{figure}
\includegraphics[width = 86mm]{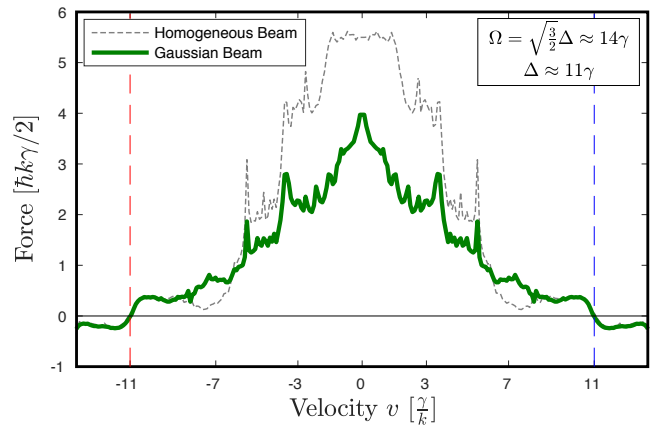}
\caption{(color online). {\bf Velocity-dependent force profile.} The parameters for argon, $\gamma = 2 \pi \times 5.87\text{\,MHz, } \Delta = 2 \pi \times 65\text{\,MHz} \approx 11 \gamma$ and $\Omega = \sqrt{\frac{3}{2}}\Delta \approx 14 \gamma$ are used for the numerical calculation (dashed). Accounting for different Rabi frequencies due to the Gaussian laser beam and therefore non-optimal bichromatic forces yields a corrected calculation (solid). The sharp resonances correspond to Doppleron resonances and the vertical dashed lines mark $|\pm kv| = \Delta$, where standard optical molasses can be seen as zero crossings of the force.}
\label{fig:nc_rabi14}
\end{figure}

The dressed-atom model gives good insight on the nature of the bichromatic force. To provide more detail, we numerically integrate the optical Bloch equations (OBE) which describe the time evolution of the Bloch vector $R = (r_{1},r_{2},r_{3})$ whose components are $r_{1} = \rho_{ge} e^{-i \omega t} + \rho_{eg} e^{i \omega t}$, $r_{2} = i(\rho_{ge} e^{-i \omega t} - \rho_{eg} e^{i \omega t})$ and $r_{3} = \rho_{ee} - \rho_{gg}$ \cite{Metcalf1999}. Using the light field configuration in \eqref{eq:doubledressedrabi}, they read
	\begin{equation} \label{eq:timeevolutionblochvector}
		\begin{aligned}
			\dot{r}_{1} =& -\Delta r_{2} - \frac{\gamma}{2} r_{1} -  4\Omega r_{3} \sin(kvt)\sin(\frac{\phi}{4})\sin(\Delta t), \\
			\dot{r}_{2} =& -\Delta r_{1} - \frac{\gamma}{2} r_{2} +  4 \Omega r_{3} \cos(kvt)\cos(\frac{\phi}{4})\cos(\Delta t), \\
			\dot{r}_{3} =& \gamma(1-r_{3}) 	+  4 \Omega [r_{1} \sin(kvt)\sin(\frac{\phi}{4})\sin(\Delta t) \\
					  &  - r_{2} \cos(kvt)\cos(\frac{\phi}{4})\cos(\Delta t)].
		\end{aligned}
	\end{equation}
The velocity dependence is included by the substitution $z=vt$ for the position. The force for a given velocity $v$ is found to be \cite{Partlow2004}
	\begin{equation}
		\begin{aligned}
			F_{\text{BC}} = 2\hbar k \Omega [ 	&-r_{1}\sin(kvt)\cos(\frac{\phi}{4})\cos(\Delta t) \\
								&+r_{2} \cos(kvt)\sin(\frac{\phi}{4})\sin(\Delta t)	].
		\end{aligned}
	\end{equation}
 In Figure \ref{fig:nc_rabi14}, the results for the case of a homogeneous light beam and for a beam with a Gaussian intensity profile are shown.

\section{Experimental Setup} 
\begin{figure}
\includegraphics[width = 86mm]{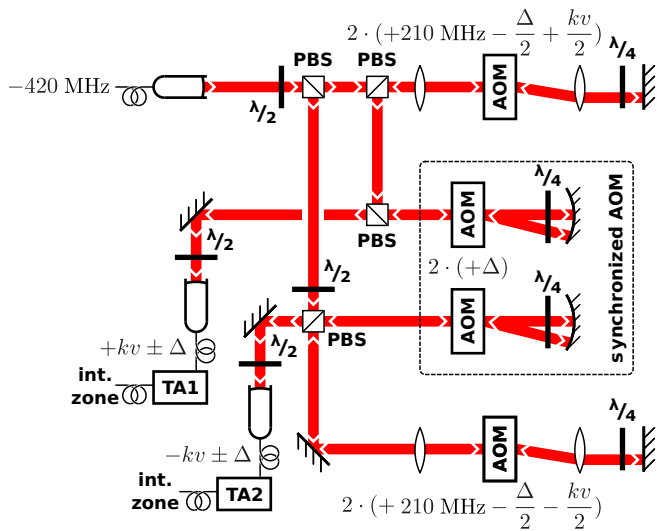}
\caption{(color online). {\bf Frequency generation.} The incoming light has a detuning of $-420$\,MHz from atomic resonance $\omega_{A}$ and is shifted upwards to $-\Delta\pm kv$ with the top and bottom AOMs. The two AOMs in between are driven by synchronized frequency generators and are adjusted for equal distribution into zeroth and first order. A relative phase shift $\phi$ can be added by tuning the relative phase of the driving rf-frequencies. This yields the four frequencies $\nu_{BC}= \omega_{A}\pm kv\pm\Delta$ which are amplified subsequently by two homebuilt tapered amplifiers.}
\label{fig:fourfrequencies}
\end{figure}

For the simplest case of atoms at rest, the bichromatic force requires two counterpropagating beams each with frequencies detuned by $\pm \Delta$ from atomic resonance $\omega_{A}$. For moving atoms with a velocity $v$ in beam direction, the Doppler shift $\delta = -kv$ has to be taken into account. This yields the four frequencies	\begin{equation}\label{eq:fourfrequencies}
		\nu_{\text{BC}} = 
		\begin{cases}
			\omega_{A}+ kv\pm\Delta & \text{for the 1}^{\text{st}}\text{ beam} \\
			\omega_{A}- kv\pm\Delta & \text{for the 2}^{\text{nd}}\text{ beam}
		\end{cases}
	\end{equation}
which are needed for the bichromatic force. Each of the two beams is intensity modulated by the beating of the contained frequencies. Together, this corresponds to the same light field configuration as in \eqref{eq:doubledressedrabi}. Figure \ref{fig:fourfrequencies} shows a schematic diagram of the setup that generates the frequencies in \eqref{eq:fourfrequencies}. The incoming light is detuned by $-420 \text{\,MHz}$ from the atomic resonance $\omega_{A}$. The outer doubly-passed acousto-optical modulators (AOM) compensate the Doppler shift for a velocity $v$ and shift the frequencies to $-\Delta -kv$ and $-\Delta +kv$ respectively, where $\delta = kv$ can be tuned by $\sim 30$\,MHz.

The second frequency component is added to the two already Doppler shifted frequencies by synchronized AOMs, which are generated by two Agilent 33250A frequency generators with rf-frequency $\Delta$. The AOMs are aligned in a way that equal distribution into the zeroth and first order takes place, resulting in generation of the four frequencies $\nu_{\text{BC}}$ in \eqref{eq:fourfrequencies} after double passing. The relative phase of the acoustic waves $\phi_{\text{AOM}}$ can be adjusted with the function generators and will directly change the relative phase $\phi$ between the zeroth order ($\omega_{A}-\Delta$) and the first order ($\omega_{A}+\Delta$) of the optical frequencies.

The generated frequencies are amplified by two homebuilt tapered amplifiers (TA) and are coupled into single-mode polarization-maintaining fibers. To avoid non-linearity in the amplification by the TAs, we constrained their outputs to $25 \text{\,mW}$ per frequency, which is a third of the maximum available power. This provides satisfying visibilities of the beating signals of $\sim$ 90\,\%.

\begin{figure}
\includegraphics[width = 86mm]{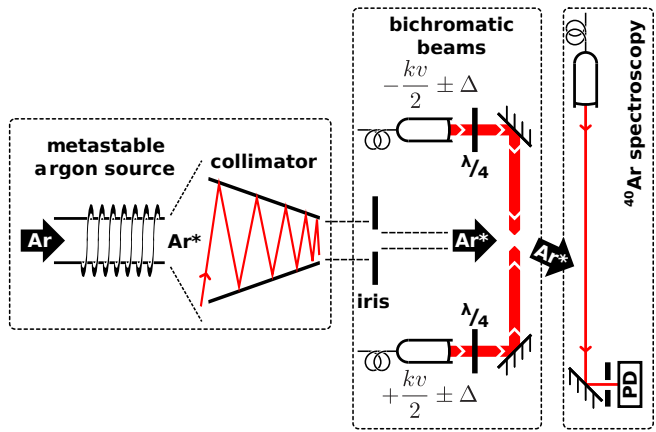}
\caption{(color online). {\bf Measurement and detection setup.} The precollimated beam of metastable $^{40}$Ar is deflected in the interaction region. The powers of the bichromatic beams have been stabilized by photodiodes placed behind the mirrors that are not shown here. The deflected argon beam aquires a Doppler shift that can be measured by an absorption spectroscopy in the detection zone.}
\label{fig:interactionzone}
\end{figure}

The states of the used $J=2 \rightarrow J'=3$ transition of $^{40}$Ar have a twelvefold degeneracy due to their different magnetic substates $m_{J}$. To obtain an effective two-level system, the light of the counterpropagating beams is circularly polarized by quarter-wave plates and is chosen to be either $\sigma^{+}$$\sigma^{+}$ or $\sigma^{-}$$\sigma^{-}$ with respect to the atoms. This optically pumps the atoms to the stretched states with maximal absolute value of $m_{J}$.

The existing experimental setup gives access to a collimated beam of metastable argon atoms \cite{Ritterbusch2014} and the force is then acting on atoms with a narrow velocity distribution of $\sim \pm 10 \text{\,m/s}$. By switching on the bichromatic interaction, an additional transverse velocity $v$ is added to the atoms. This is measured by an absorption spectroscopy (see Fig.\,\ref{fig:interactionzone}). Compared to the case of a non-deflected beam without additional force $F$, the deflected atoms acquire a Doppler shift
	\begin{equation} \label{eq:dopplershift}
		\delta = -kv = -ka\cdot t_{\text{int}} = -\frac{kF}{m} \cdot \frac{d}{v_{\text{long}}}
	\end{equation}
with interaction time $t_{\text{int}}$, light beam diameter $d$ and the longitudinal velocity $v_{\text{long}}$ of the atoms perpendicular to the force. The spectroscopy light is generated by changing an offset-locked frequency from the existing laser system. This provides convenient access to an absolute frequency scale.

\section{Results and Discussion} 
All results are given relative to the maximal spontaneous scattering force $F_{\text{sp}}$ in units of $\hbar k \gamma/2$, which was measured by blocking one beam and ensuring the saturation of the scattering force induced by the remaining modulated beam. This technique allows for a direct comparison of the methods, since it obliterates the dependence on the interaction length. With a detuning of $\Delta = 2\pi \times 65 \text{\,MHz}$, a bichromatic force three times as high as $F_{\text{sp}}$ has been achieved. 

A correction to the numerical model has to be made since the interaction zone consists of counterpropagating beams with Gaussian intensity distributions. Therefore, the Rabi frequency is not constant and the optimal condition is not fulfilled during the whole interaction time, leading to significantly lower bichromatic forces.

Still, the measured values are approximately 20\,\% lower than the theoretical expectation. One explanation for this discrepancy is the multilevel structure of the transition, due to which the stretched states, which correspond to a two-level system, are only reached after optical pumping. The bichromatic force can only exert its full strength in this two-level scenario. To take this approximation into account, the following numerically calculated curves are scaled with a factor of 0.8. The errors of the measured data arise from the $1\sigma$ region of the absorption spectroscopy fit.

\subsection{Phase dependence}
\begin{figure}
\includegraphics[width = 86mm]{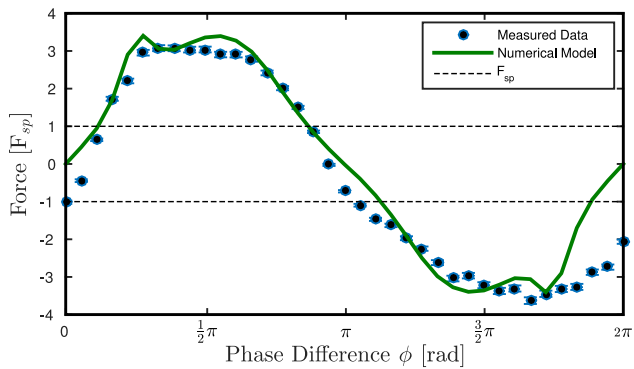}
\caption{(color online). {\bf Phase dependence of the force.} The bichromatic force has been measured with absorption spectroscopy for different relative phases $\phi$ of the light fields. The optical phase can be controlled arbitrarily by tuning the phases of the sine waves on the acousto-optical modulators. Thereby, the direction and strength of the force is changed in agreement with theoretical predictions which are shown by the solid curve. As expected from the doubly-dressed atom model, the force vanishes near phases $\phi = 0$, $\phi = \pi$ and $\phi = 2\pi$.}
\label{fig:phasedependence}
\end{figure}

Characteristic for the bichromatic force is its dependence on the phase difference $\phi$ between the two counterpropagating laser beams \cite{Metcalf1999}. It can be tuned conveniently by changing the relative phase of the two rf-frequencies on the acousto-optical modulators. Figure \ref{fig:phasedependence} shows this effect up to a complete inversion of the force. It is in good agreement with the theoretical expectation.

\subsection{Dependence on Rabi frequency}
\begin{figure}
\includegraphics[width = 86mm]{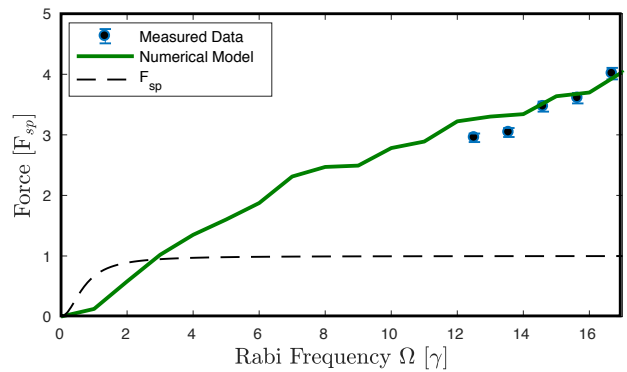}
\caption{(color online). {\bf Rabi frequency dependence of the force.} The bichromatic force has been measured for different intensities which are mapped to corresponding Rabi frequencies by using $\Omega = \gamma \sqrt{\frac{I}{2 I_{S}}}$ and $I_{S} = 14.4 \text{ W}/\text{m}^2$ for circularly polarized light. The bandwidth of the setup has allowed Rabi frequencies roughly between $12\gamma$ to $17\gamma$ since the detuning has to be scaled with $\Delta = \sqrt{\frac{2}{3}}\Omega$.  The measured data points show a gain of the force, as expected from the doubly-dressed atom model and the solid curve shows the numerically calculated force. The spontaneous scattering force (dashed) increases faster but saturates eventually.}
\label{fig:detuningdependence}
\end{figure}

The presented theoretical models all predict a boost of the force by increasing the Rabi frequency $\Omega$ while changing the detuning $\Delta$ accordingly. Due to the limited AOM bandwidth and optical power, a detuning range of $60-80 \text{\,MHz}$ could be achieved with the setup, which corresponds to attainable Rabi frequencies of about $12\gamma$ to $17\gamma$. Figure \ref{fig:detuningdependence} shows the results. The force $F_{\text{BC}}$ increases linearly with $\Omega$ and the depicted numerical solution is in good agreement. Therefore, even stronger bichromatic forces should be accessible with higher Rabi frequencies.

\subsection{Detuning width of the force}
\begin{figure}
\includegraphics[width = 86mm]{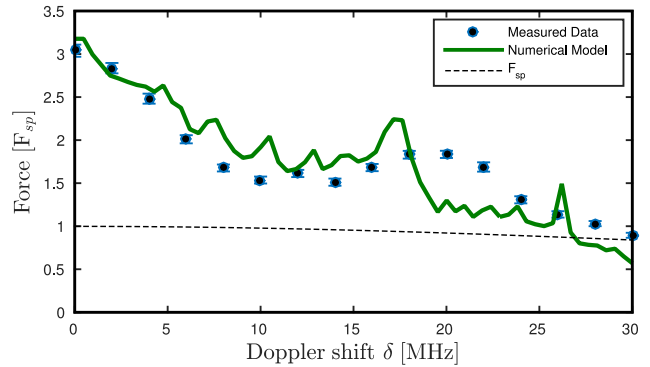}
\caption{(color online). {\bf Detuning width of the force.} The bichromatic force profile has been measured for different frequency offsets which correspond to Doppler shifts $\delta = |kv|$.  The data points resemble the rough structure of the force profiles that have been calculated with the optical Bloch equations, shown by the solid curve. These values are further corrected by accounting for the inhomogeneous Gaussian intensity distribution. The dashed line is the expected spontaneous scattering force with the same intensity and is saturated over most of the shown range.}
\label{fig:dopplershiftdependence}
\end{figure}

For cooling purposes, the Doppler shift of the atoms plays a major role and the dependence on the induced detuning $\delta = \pm kv$ from atomic resonance is of interest. In the given setup, the velocity of the atoms in light beam direction is close to zero. To investigate the behaviour of a finite Doppler shift, additional detunings of $\delta = + kv$ and $\delta = - kv$ have been added to the bichromatic beams so that atoms will appear Doppler shifted in a reference frame moving with $v$. Figure \ref{fig:dopplershiftdependence} shows the measured force for different $\delta = |kv|$ in the range between $0$ and $30 \text{\,MHz}$, which is limited by the given experimental setup.

\begin{figure*}
\includegraphics[width = 178mm]{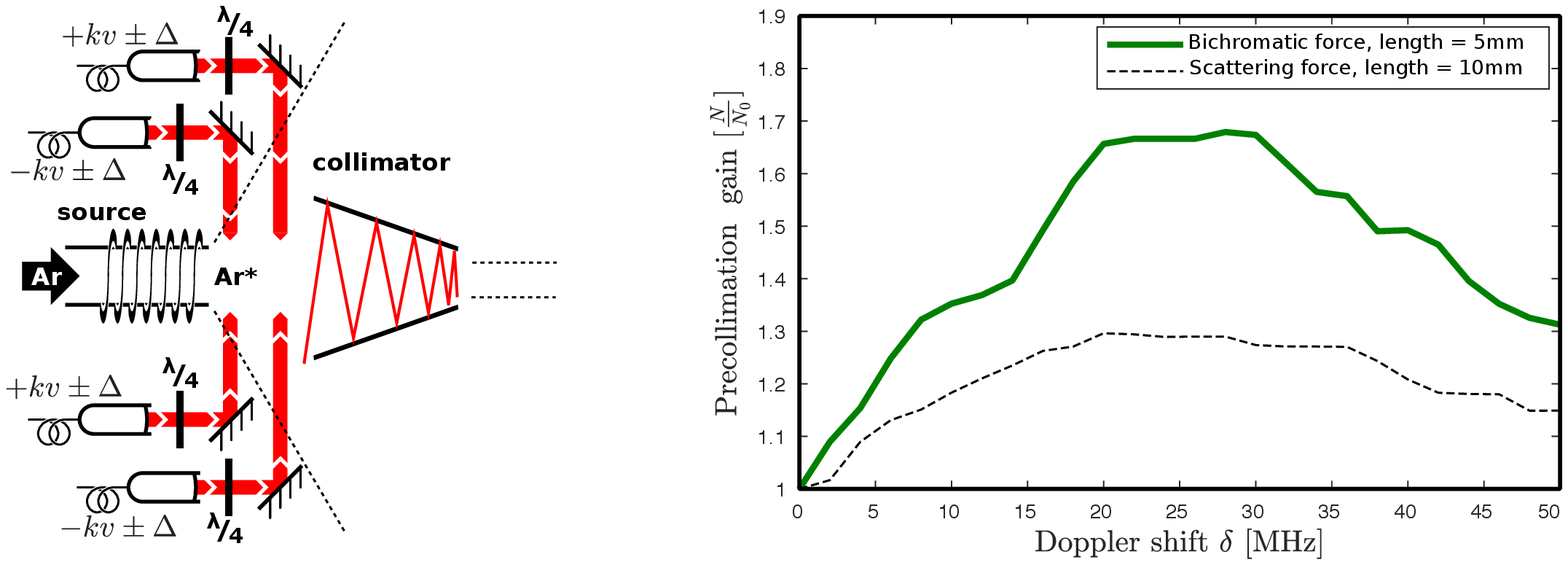}
\caption{(color online). {\bf Proposed setup for a bichromatic transverse kicker.} The transversal velocity capture range of the existing collimator covers only $\sim$ 6\,\% of the atoms leaving the effusive source. This can be increased by a transverse kicker, a precollimation stage to decelerate faster atoms below the critical velocity of the collimator (schematic diagram shown left). The gain as a function of laser detuning (right) to compensate the Doppler shift $\delta$ is shown for bichromatic forces with two separate interaction regions of 5 mm length (solid) and spontaneous scattering forces with a 10 mm long interaction region (dashed). The limited interaction time favors a bichromatic transverse kicker, even if the interaction region is halved for every direction due to independent collimation of the two perpendicular dimensions.}
\label{fig:precollimation}
\end{figure*}

Under optimal conditions, the width (FWHM) of the bichromatic force is approximately the detuning $\Delta$ \cite{Williams1999}, which can also be estimated from Figure \ref{fig:nc_rabi14}. The measured data show a comparable width of the distribution and are also in good agreement with the numerical results. \\

\section{Conclusion and Outlook}

This work presents a first study of the bichromatic force on argon where many parameters are chosen for experimental convenience. For practical use, the strength of the force is only limited by the detuning $\Delta$ and therefore by the available optical power to fulfill the condition for the optimal Rabi frequency. The key signatures, which are the magnitude and dependence on the relative phase difference $\phi$, were observed and demonstrate the bichromatic force on metastable $^{40}$Ar. The numerical model using optical Bloch equations included the Gaussian intensity distribution of the light beams and yields comparable behaviours.

The main purpose for investigating the bichromatic force for ATTA is a more efficient collimation of the atoms leaving the source. In comparison with usual optical molasses, this requires four times the interaction length since each direction has to be set up independently. However, it is possible to overlap two pairs of bichromatic beams which would halven the required interaction length. In the same dimension (e.g. horizontal or vertical), the three frequencies $\omega_{A}$ and $\omega_{A}\pm 2\Delta$ lead to a bichromatic force from both directions with center at $\delta = \pm kv = \pm \Delta$. It is also possible to overlap the interaction region of perpendicular beams by adjusting the relative phase difference of each frequency to $\phi = \frac{\pi}{2}$, which requires four synchronized frequencies.

Using the results obtained in this work, the velocity distribution of the effusive source and the capture range of the collimator, an estimation for the gain of a bichromatic precollimator (transverse kicker) yields a factor of $\sim 1.7$ per dimension (see Fig.\,\ref{fig:precollimation}). Due to the reasons given in the previous paragraph, the interaction region is chosen only half as long as for the case of the scattering force. For the application with $^{39}$Ar, an additional challenge arises due to its hyperfine splitting, which needs to be addressed by proper optical repumping. Besides enhancing the performance of ATTA, other experiments that rely on high flux will also benefit from such a transversal cooling scheme. Examples are the trapping of unstable Ne isotopes for addressing questions in high energy physics \cite{Ohayon2015} or measurements of physical constants as the rms nuclear charge radii of $^{6}$He and $^{8}$He \cite{Mueller2007}.

\section*{Acknowledgements}
We thank Helmut Strobel for many discussions and careful proofreading of the manuscript. We thank Harold Metcalf and Chris Corder for discussions. This work was supported by the Deutsche Forschungsgemeinschaft (DFG, German Research Foundation) and the European Research Council (ERC) under the European Union's Horizon 2020 research and innovation programme (Grant agreement No 694561). S. E. acknowledges support from the Studienstiftung des deutschen Volkes.
  

\bibliographystyle{APS}

%

\end{document}